\documentclass[conference]{IEEEtran}
\usepackage[utf8]{inputenc}
\usepackage[T1]{fontenc}
\usepackage{graphicx}
\usepackage{grffile}
\usepackage{longtable}
\usepackage{wrapfig}
\usepackage{rotating}
\usepackage[normalem]{ulem}
\usepackage{amsmath}
\usepackage{textcomp}
\usepackage{amssymb}
\usepackage{capt-of}
\usepackage{balance}
\usepackage{units}
\usepackage{cite}
\usepackage{tabularx}
\usepackage{multirow}
\usepackage{comment}
\usepackage[table]{xcolor}
\usepackage[nolist]{acronym}
\usepackage{pgfplots}
\usepackage{dblfloatfix}
\usepackage{tikz}

\begin{acronym}[ACRONYM]
\acro{AI}{artificial intelligence}
\acro{AoA}{angle-of-arrival}
\acro{AoD}{angle-of-departure}
\acro{BS}{base station}
\acro{D-MIMO}{distributed multiple-input multiple-output}
\acro{DFTS}{discrete Fourier transform spread}
\acro{EM}{electromagnetic}
\acro{E2E}{end-to-end}
\acro{GDOP}{geometric dilution of precision}
\acro{GNSS}{global navigation satellite system}
\acro{IQ}{in-phase and quadrature}
\acro{JCS}{Joint Communication and Sensing}
\acro{JRC}{joint radar and communication}
\acro{JRC2LS}{joint radar communication, computation, localization, and sensing}
\acro{ICI}{inter-carrier interference}
\acro{IOO}{indoor open office}
\acro{IoT}{Internet of Things}
\acro{IN}{infrastructure node}
\acro{KPI}{key performance indicator}
\acro{LoS}{line-of-sight}
\acro{MIMO}{multiple-input multiple-output}
\acro{mmWave}{millimeter-wave}
\acro{NLoS}{non-line-of-sight}
\acro{NR}{new radio}
\acro{OFDM}{orthogonal frequency-division multiplexing}
\acro{OTFS}{orthogonal time-frequency-space}
\acro{PRS}{positioning reference signal}
\acro{QoS}{Quality of Service}
\acro{RAN}{radio access network}
\acro{RAT}{radio access technology}
\acro{RedCap}{reduced capacity}
\acro{RF}{radio frequency}
\acro{RIS}{reconfigurable intelligent surface}
\acro{RTK}{real-time kinematic}
\acro{RTT}{round-trip-time}
\acro{SLAM}{simultaneous localization and mapping}
\acro{SNR}{signal-to-noise ratio}
\acro{ToA}{time-of-arrival}
\acro{TDoA}{time-difference-of-arrival}
\acro{TR}{time-reversal}
\acro{TXRX}[Tx/Rx]{transmitter/receiver}
\acro{TX}[Tx]{transmitter}
\acro{RX}[Rx]{receiver}
\acro{UE}{user equipment}
\end{acronym}

\begin{document}
\bstctlcite{IEEEexample:BSTcontrol}

\title{6G Radio Requirements to Support Integrated Communication, Localization, and Sensing}

\author{
Henk Wymeersch\IEEEauthorrefmark{7}, 
Aarno Pärssinen\IEEEauthorrefmark{2}, 
Traian E. Abrudan\IEEEauthorrefmark{8}, 
Andreas Wolfgang\IEEEauthorrefmark{4},
Katsuyuki Haneda\IEEEauthorrefmark{6},\\  
Muris Sarajlic\IEEEauthorrefmark{1}, 
Marko E. Leinonen\IEEEauthorrefmark{2},
Musa Furkan Keskin\IEEEauthorrefmark{7}, 
Hui Chen\IEEEauthorrefmark{7},  \\
Simon Lindberg\IEEEauthorrefmark{4},
Pekka Ky\"{o}sti\IEEEauthorrefmark{2}, 
Tommy Svensson\IEEEauthorrefmark{7}, 
Xinxin Yang\IEEEauthorrefmark{4},
\\ 
 \IEEEauthorrefmark{7}Chalmers University of Technology, Sweden
 \IEEEauthorrefmark{1}Ericsson Research, Sweden\\
 \IEEEauthorrefmark{4}Qamcom Research and Technology, Sweden
  
 \IEEEauthorrefmark{2}University of Oulu, Finland\\
 \IEEEauthorrefmark{6}Aalto University, Finland
 \IEEEauthorrefmark{8}Nokia Bell Labs, Finland 
}

\maketitle

\begin{abstract}
6G will be characterized by 
extreme use cases, not only for communication, but also for localization, and sensing. The use cases can be directly mapped to requirements in terms of standard \acp{KPI}, such as data rate, latency, or localization accuracy. The goal of this paper is to go one step further and map these standard \acp{KPI} to requirements on signals, on hardware architectures, and on deployments. Based on this, system solutions can be identified that can support several use cases simultaneously. Since there are several ways to meet the \acp{KPI}, there is no unique solution and preferable configurations will be discussed. 
\end{abstract}
\acresetall
\section{Introduction}
\label{sec:intro}
6G will support extreme use cases, requiring gigabit-per-second peak data rates with millisecond \ac{E2E} latency \cite{uusitalo20216g}. 
Secondly, 6G will rely on a variety of new enablers at the physical layer \cite{giordani2020toward}, including the utilization of frequency bands in the upper \ac{mmWave} range ($\unit[100\text{--}300]{GHz}$), novel \ac{RF} architectures, exploration of hardware-friendly, energy-efficient waveforms and beamforming, and the use of distributed large MIMO system and \acp{RIS}. 
Finally, 6G will feature a tight integration among communication, localization, and sensing~\cite{de2021convergent,wymeersch2021integration}. 
Integration will not only enable exciting, but challenging, new use cases, 
but also improve communication functionality 
 as well as other services
 \cite{Hexa-X-D3.1}. 
Use case requirements are generally specified in terms of \acp{KPI}, which
in turn impose requirements on the signals (e.g., bandwidth, waveform, and modulation) \cite{tataria20216g}, the hardware architectures (e.g., carrier, channelization, array type, output power) \cite{boulogeorgos2018terahertz}, and deployments (e.g., placement of (distributed) \acp{BS} and \acp{RIS}) \cite{strinati2021reconfigurable}. 

In this paper, we investigate the requirements on signals, hardware architectures, and deployments for 6G integrated communication, localization, and sensing. Starting from several 6G use cases, we state expected \acp{KPI}, list the possible options in terms of signals, hardware architectures, and deployments, followed by discussing how they should be determined to support the considered use cases. As there is no unique solution to this design problem, 
several alternatives are discussed, with the aim of finding commonalities and possibly a joint design that can serve several 6G use cases simultaneously. The main contribution of this paper is to provide both a methodology for determining such requirements, as well as listing initial solutions. We also highlight several of the (sometimes subtle) synergies and trade-offs that should be considered in overall 6G radio.  

In terms of structure, this paper starts with a recap of 6G use cases and KPIs (Section \ref{sec:use-cases}), followed by a brief overview of the components of the radio channel (antennas, RF technology, and wave propagation) at upper mmWave frequencies (Section \ref{sec:RFtech}). Section \ref{sec:indreq} describes the identified degrees of freedom in terms of  signals, hardware architecture, and deployments. Then, in Sections \ref{sec:Signals}--\ref{sec:deployments}, the specific requirements are determined, 
needed to support the use cases from Section \ref{sec:use-cases}. We conclude the paper with Section \ref{sec:conclusion}.

\begin{figure}
    \centering
    \includegraphics[width=0.7\columnwidth]{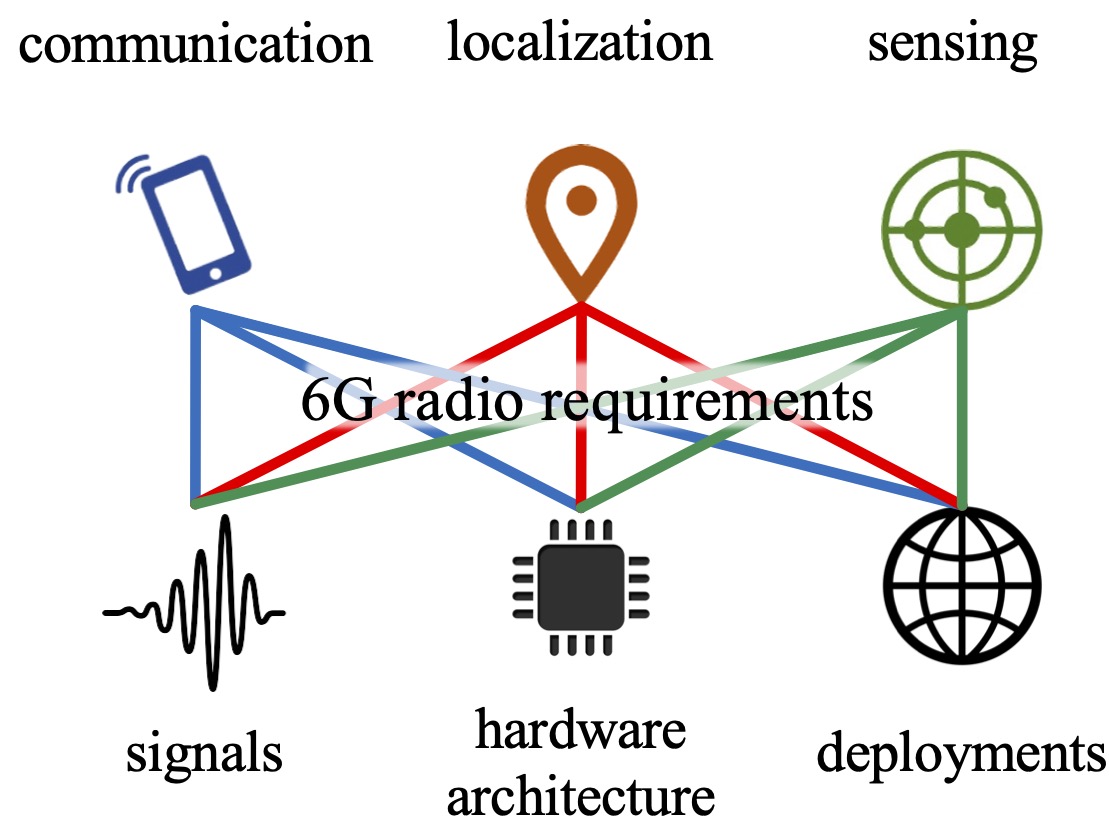}
    \caption{Communication, localization, and sensing each impose requirement on signal, hardware architectures, and deployments, which should be met by 6G radio. }
    \label{fig:overviewV2}
    \vspace{-6mm}
\end{figure}

\section{6G Use Cases and \acp{KPI}}
\label{sec:use-cases}
In this section, we recap several 6G use cases and list their 
\acp{KPI}, for all three considered applications: communication (C1--C2), localization (L1--L3), and sensing (S1--S2).

\subsection{Communication}
Following \cite{Hexa-X-D2.1}, we propose two communication system design scenarios corresponding to 6G use-cases with advanced design requirements.
    \emph{Very short-range wireless access (C1)} $\unit[100]{Gbps}$ per-user rate, $\unit[0.1\text{--}1]{ms}$ \ac{E2E} latency, $\unit[10]{m}$ link range.
    \emph{Short-range wireless access (C2)} $\unit[10]{Gbps}$ per-user rate, $\unit[{<}1]{ms}$ \ac{E2E} latency, $\unit[100]{m}$ link range. 

\subsection{Localization}
Following \cite{saily2021positioning}, we divide localization scenarios into three categories, with KPIs derived from \cite{Hexa-X-D3.1}.
    \emph{High-accuracy positioning (L1)} 
for very fine maneuvers or coordination. An example in this category is \emph{augmented reality}, with $\unit[1]{cm}$ location accuracy, $\unit[1]{^{\circ} }$ orientation accuracy, $\unit[100]{Hz}$ update rate, and $\unit[10]{m}$ link range.
\emph{Low-latency positioning (L2)}, for example for 
\emph{collaborative robot localization}, requiring $\unit[10]{cm}$ location accuracy, $\unit[1]{^{\circ} }$ orientation accuracy, $\unit[1]{kHz}$ update rate, and $\unit[30]{m}$ link range. 
\emph{Low-complexity positioning (L3)} for \ac{IoT} localization with \ac{RedCap} devices. An example is \emph{remote sensing}, with $\unit[1\text{--}10]{m}$ localization accuracy, $\unit[1]{Hz}$ or lower update rate, and more than $\unit[1]{km}$ link range.

\subsection{Sensing}
A natural breakdown is into monostatic and bi-/multi-static sensing. 
 \emph{Monostatic sensing (S1)} 
    involves a transmitter and receiver are co-located, sharing common knowledge of the data and have a shared clock, and 
provides radar-like mapping (e.g., for automotive applications), with $\unit[10]{cm}$ distance accuracy, $\unit[0.04]{m/s}$ velocity accuracy, $\unit[3]{^{\circ} }$ angular resolution, $\unit[0.2]{^{\circ} }$ angular accuracy, maximum range of $\unit[50]{m}$, and $\unit[25]{Hz}$ update rate \cite[Table C2]{5GCAR-D2.1}. This application is similar to the \emph{sensor infrastructure web} use case from \cite{Hexa-X-D3.1}. We only consider monostatic sensing with communication signals, not dedicated sensing waveforms.
\emph{Bi-/multi-static sensing (S2)} involves 
physically separate \acp{TX} and \acp{RX}, e.g., for \emph{robotic object sensing}, with $\unit[1]{cm}$ localization  resolution, as well as angular resolution below $\unit[1]{^{\circ}}$, $\unit[0.1]{m/s}$ velocity resolution, and up to $\unit[1]{kHz}$ update rate.
%

\section{Radio Channel and RF Technologies}
\label{sec:RFtech}
The radio channel is composed of antennas/arrays connected to RF transceivers at ends of the link, with the  multipath wave propagation channel between them. Combined, they 
determine the practical link performance within the constraints imposed by the wave propagation and semiconductor/material physics. The following summarizes some viewpoints about them, focusing on upper \ac{mmWave} frequencies.

\subsection{Antennas and Arrays}
At both
lower  and 
upper \ac{mmWave}, most likely 
hybrid architectures of analog RF and digital baseband beamforming will be used to support a massive number of antenna elements in the array. Due to challenges in integrated designs of an antenna array and its feeding network for RF phase shifting, 
practical implementation of an integrated phased antenna array may need a compromise in a beam scanning range and some extents of grating lobes. 
Lens antennas and radomes, despite  bulkiness, allow improvement of the beam scan range of antenna arrays, e.g.,~\cite{Benini18_TAP}, and are applicable to upper \ac{mmWave} RF.
The knowledge of radiation patterns of antennas and arrays is essential in localization and sensing, while it is less important in communications but would enhance the link performance in, e.g., beam searching. Antenna and array {calibration} methods,  in an anechoic chamber or over-the-air, would be an important technical element~\cite{Pohlmann21_EuCAP}. 

\subsection{RF Technology}
Array performance is dependent on the amplifier performance both in transmit and receive side. The former defines the largest possible output power while the latter minimum noise figure. Both are to certain extent bandwidth dependent and even more carrier frequency dependent. Fundamental limits stem from the properties of semiconductor technologies and wiring losses. Silicon and compound semiconductors including different transistor types, sizes and technology nodes determine maximum operating frequency as maximum unilateral gain ($f_{\max}$) or transition frequency. As carrier frequency gets higher, the gain, fundamental requirement of RF processing, approaches unity while available output power decreases and noise increases rapidly. Recent example of a state-of-the-art $\unit[290]{GHz}$ amplifier in SiGe BJT process demonstrates more than $\unit[10]{dB}$ of gain at $2/3$ of $f_{\max}$ at the cost of significantly lowered dynamic range compared to a low frequency counterpart using the same technology~\cite{Singh21}. RF transceivers for single link have been demonstrated at frequencies above $\unit[200]{GHz}$ for short range ($\unit[1]{m}$) communications trials achieving $\unit[100]{Gbps}$~\cite{Pfeiffer20}. 

\subsection{Wave Propagation}
The most important characteristic of wave propagation is pathloss, with 
upper \ac{mmWave} RFs showing pathloss exponents of {\it omni-directional channels} around $2$ and $3$ in outdoor cellular \ac{LoS} and \ac{NLoS} environments~\cite{Xing21_ICC}. Even though diffraction coefficients are smaller as the RF increases, reflections on concrete walls, metal lampposts and tinted glasses can deliver power from one link end to another, making the link connectivity in \ac{NLoS} feasible through one or multiple reflections even for upper \ac{mmWave}. The finding of pathloss exponents does not differ from available insights for lower \ac{mmWave} RF according to the comparison of indoor hotspot channels~\cite{Nguyen21_VTCS}.
The number of multipaths or clusters is an important degree of freedom. 
More clusters indicate increased possibility of spatial multiplexing to send different data sets over possibly orthogonal beams. 
 Existence of multiple clusters also implies the possibility of device localization and sensing through them. Measurements show the number of spatial clusters to be $0.69$ and $1.82$ for \ac{LoS} and \ac{NLoS} urban microcellular links~\cite{Ju21_Globecom} while channels support one to four beams in an indoor entrance hall~\cite{Kyosti21_AWPL}. 

\begin{table*}[!b]
  \centering
  \rowcolors{3}{gray!10}{white}
  \resizebox{0.99\textwidth}{!} {
\begin{tabular}{|p{25mm}|p{20mm}|p{20mm}|p{20mm}|p{20mm}|p{20mm}|p{20mm}|p{20mm}|p{20mm}}
\hline 
\rowcolor{gray!50} Use case  & C1  & C2    & L1  & L2 & L3  & S1  & S2 \tabularnewline
\hline 
\hline 
\textbf{Signals}  &    &  &  &  &  &    & \tabularnewline
Bandwidth  
& (a) $\unit[4]{GHz}$ \newline  (b) $\unit[10]{GHz}$  
& (a) $\unit[0.4]{GHz}$ \newline (b) $\unit[5]{GHz}$ 
& (a) $\unit[2\text{--}4]{GHz}$ \newline (b) $\unit[500]{MHz}$  
& $\unit[0.5\text{--}1]{GHz}$ 
& $\unit[{<}500]{MHz}$  
& (a) $\unit[2\text{--}4]{GHz}$ \newline (b) $\unit[{<}1]{GHz}$
& $\unit[2\text{--}4]{GHz}$ 
\tabularnewline
Waveform  
& any 
& any 
& (DFTS-)OFDM   
& (DFTS-)OFDM 
& any  
& (DFTS-)OFDM   
& (DFTS-)OFDM 
\tabularnewline
Modulation  & coherent or \newline  non-coherent & coherent or \newline  non-coherent &    coherent  & coherent  & coherent or non-coherent   &  coherent  & coherent
\tabularnewline
Signal shaping  & space in \newline some scenarios  &  space and/or freq. \newline in some scenarios  &    freq., time, and space   & freq.~and space  & no  &  freq., time, and space   & trade-off with comm.  
\tabularnewline
\hline 
\textbf{Hardware Arch.}  &  &  &    &  &  &    & \tabularnewline
Carrier* & $\unit[60\text{-}140]{GHz}$   & (a) $\unit[\text{sub-}6]{GHz}$ 
\newline (b) $\unit[60 \text{-} 140]{GHz}$ 
& (a) $<\unit[30]{GHz}$ \newline (b) $\unit[60 \text{-} 140]{GHz}$ &$\unit[60 \text{-} 140]{GHz}$ & $\unit[6\text{--}30]{GHz}$ & $\unit[60 \text{-} 140]{GHz}$  & $\unit[60 \text{-} 140]{GHz}$ \tabularnewline
Channelization & optional  & optional   &     no  &  no  &  no  & not preferred   & not preferred \tabularnewline

\acs{IN} array type  & analog or hybrid  & analog or hybrid  &     analog or hybrid & hybrid or digital  & analog   & analog or hybrid   & analog or hybrid \tabularnewline
\acs{UE} array type  &  analog or hybrid & analog or hybrid  &     analog or hybrid & analog or hybrid  & SNR boost   & analog or hybrid   & hybrid or digital  \tabularnewline
\acs{IN} array size (per dim.) & 10-20  & 10-20  &    (a) 10--20 \newline (b) 50--100  & 10--20   & SNR boost  & (a) 10--20 \newline (b) 40--50   & (a) 10--20 \newline (b) 40--50  \tabularnewline
\acs{UE} array size (per dim.) & 4-8  & 4-8  &    10--20   & 10--20 &  SNR boost & (a) 100 \newline (b) 10--20    & 100 \tabularnewline
Transmit power  & medium  & medium  &     low (see \eqref{eq:SPEB}) & higher \newline (small $T$ in  \eqref{eq:SPEB})  & higher \newline   (large $d$ in  \eqref{eq:SPEB}) &  high (high path loss)  & high (high path loss) \tabularnewline
\hline 
\textbf{Deployments}  &  &  &    &  &  &    & \tabularnewline
Placement around each  device 
& (a) $\geq 1$ \acs{IN}  multi-stream or D-MIMO
\newline (b) $\geq 1$ \acs{IN}  single stream
& (a) $\geq 1$ \acs{IN}  multi-stream or D-MIMO 
\newline (b) $\geq 1$ \acs{IN}  single stream
& $\geq 4$ \acsp{IN} in \acs{LoS} (\acs{TDoA}, 3D), 
    \newline $\geq 2$ \acsp{IN} in \acs{LoS} (\acs{AoA}, 3D)
& same as L1
& $\geq 4$ \acsp{IN} in \acs{LoS} (\acs{TDoA}, 3D) 
& N/A 
& (a) $\geq 1$ \acs{RX} \acsp{IN} (TDoA + \acs{AoA} + AoD), Tx-Rx in \acs{LoS}
\newline (b) $\geq 1$ \acs{RX} \acsp{IN} (\acs{ToA} + \acs{AoA} + AoD), Tx-Rx in \acs{NLoS}
\tabularnewline
Synchronization   
& $\unit[1\text{--}10]{ns}$ (\acs{D-MIMO}) 
& $\unit[1\text{--}10]{ns}$  (\acs{D-MIMO})  
& $\unit[100]{ps}$ (\acs{TDoA}) 
& $\unit[0.5]{ns}$ (\acs{TDoA}) 
& $\unit[1\text{--}10]{ns}$ (\acs{TDoA})   
& N/A  
& (a) N/A \newline (b) $\unit[100]{ps}$ 
\tabularnewline
\acs{IN} knowledge  
& position: area-level
& position: area-level
& position: mm-level
    \newline orientation: $\leq 0.1^{\circ}$ (\acs{AoA})
& position: cm-level,
    \newline orientation: $\leq 1^{\circ}$ (\acs{AoA})
& position: m-level,
\newline orientation: $\leq 1^{\circ}$ (\acs{AoA})
& N/A
& position: mm-level,
    \newline orientation: $\leq 0.5^{\circ}$ (\acs{AoA})
\tabularnewline
\hline 
\end{tabular}}
  \vspace{5mm}
  \caption{Summarizing table on the requirements on signals, hardware architectures, and deployments. Acronyms are defined in the text. {*Caution: numbers are in reality technology-dependent, with noise figure and maximum transmit power being frequency-dependent.}}
  \label{tab:SummarizingTable}
  \vspace{-5mm}
\end{table*}

\section{Signals, Architectures, and Deployments}
\label{sec:indreq}
\vspace{-0mm}
We describe the degrees of freedom in terms of signals, hardware architectures, and deployments. These will then be elaborated in the subsequent sections to support the use cases from Section \ref{sec:use-cases}, and summarized in Table \ref{tab:SummarizingTable}. 

\subsection{Signals}
The radio signals are considered to have the following 4 degrees of freedom: the \emph{aggregate bandwidth}\footnote{The aggregate bandwidth does not have to be contiguous in frequency and can be fragmented into multiple RF chains with separate smaller-bandwidth ADCs. For full utilization of the entire bandwidth, phase coherency should be maintained across the different chains.} (in Hz), the \emph{waveform type} (examples are \ac{OFDM} \cite{RadCom_Proc_IEEE_2011}, \ac{DFTS} \ac{OFDM} \cite{dfts_ofdm_commag,dfts_ofdm_thz_isac}, \ac{OTFS} \cite{OTFS_RadCom_TWC_2020}, single carrier \cite{80211_Radar_TVT_2018}), the \emph{modulation type} (e.g., constant-modulus or QAM, non-coherent or coherent),  the \emph{signal shaping} (in time-frequency-space domains, including precoding, combining, pilot allocation, and power allocation).

\subsection{Hardware Architectures} 
Hardware specifications originate from signal quality and range requirements. The former sets bandwidth and \ac{SNR}. 
On the other hand, underlying technologies determine minimum achievable noise and maximum transmitted power that are physics, technology, bandwidth, and carrier frequency dependent. In addition, also energy consumption and cost  need to be considered carefully if we need to parallelize the signal paths. Wider bandwidth leads inevitably to higher noise and  to the need to search for available radio spectrum at higher carrier frequencies. Both are impacting negatively on achievable performance. Taking these into account we have some degrees of freedom in frequency to arrange signals in \emph{carrier, and channelization}  domains~\cite{channelization_2017}. In the spatial domain, we can increase the range by beamforming or the rate with MIMO, 
either in centralized or distributed manner. Depending on the targets we have the choice of \emph{array type} (e.g., analog, hybrid, planar), \emph{array size} (number of elements). 
RF phased arrays already exist in 5G NR operating at lower \ac{mmWave} region, which are needed even in mobile equipment for decent range~\cite{Dunworth18}. 
Careful link analysis for line-of-sight and other anticipated radio channel conditions is needed to balance all the requirements with hardware properties including antennas to evaluate the feasibility of the radio transceiver hardware for different deployments\cite{Rikkinen20}. Hence, hardware architecture in digital systems needs to determine the number of parallel digitized radio channels to achieve sufficient bandwidth and then channelize it into different radio paths in frequency and spatial domain while utilizing array gain for compensating radio path loss of every available beam\cite{Kyosti21_AWPL}. 

\subsection{Deployments} The deployments are largely limited by cost and include the following 3 degrees of freedom: the \emph{placement of \acp{IN}}, including fixed and mobile base stations and \acp{RIS} (number of \acp{IN} and their positions and orientations), the \emph{level of synchronization between \acp{IN}} (time synchronization and phase synchronization), the \emph{level of knowledge needed regarding the \ac{IN} configuration} (e.g., location, orientation, phase center). In distributed deployments, we can split the burden to several physically separated radio units with additional cost.

\section{Requirements on Signals} \label{sec:Signals}
\begin{figure}
    \centering
    \input{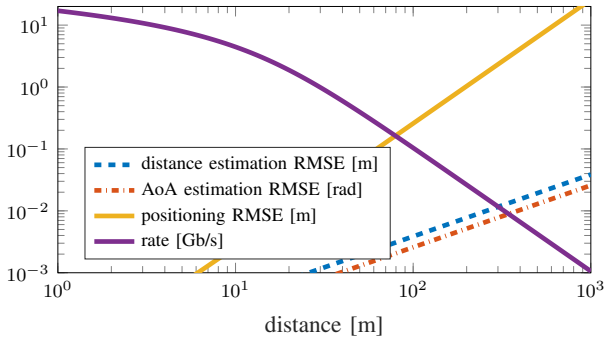}
    \caption{Example showing how the ability to estimate distance \ac{AoA} is affected by the distance between user and base station, for an uplink scenario according to \cite[Annex C]{Hexa-X-D3.1} using 120 \ac{OFDM} symbols for localization (lasting about 125 us). }
    \label{fig:tradeoff}
    \vspace{-3mm}
\end{figure}
To understand the involved trade-offs, Fig.~\ref{fig:tradeoff} shows the performance of time-based distance, \acf{AoA} estimation, and positioning in an uplink scenario under resolved \ac{LoS} with respect to the distance, following the parameters from \cite[Annex C]{Hexa-X-D3.1}, with a $\unit[140]{GHz}$ carrier, $\unit[2]{GHz}$ bandwidth, and $\unit[14]{dBm}$ transmit power. We see that even though distance and \ac{AoA} can be well estimated at far distances thanks to the large integrated SNR, positioning (combining distance and angle) quickly degrades. The rate,  assuming perfect beam alignment at the BS, degrades with distance. 

\subsection{Communication}
Assuming single-stream spectrum efficiencies of $\unit[2\text{--}6]{bps/Hz}$ and $1\text{--}4$ parallel spatial streams (depending on the array and spatial richness of the channel), the \emph{aggregate bandwidth} requirement for supporting C1 is $\unit[4\text{--}10]{GHz}$, whereas it is $\unit[0.4\text{--}5]{GHz}$ for C2. The choice of \emph{waveform type} and \emph{modulation} is largely dependent on the choice of hardware architecture, but in general waveforms with low envelope variations and inherent robustness to RF impairments are preferred. Waveforms based on OFDM have benefits in terms of backward compatibility and adaptability. 
Finally, if multi-stream transmission is to be employed, beamspace \emph{signal shaping} is required for both C1 and C2; furthermore, flexible frequency-domain multi-user allocation may be of importance for C2, whereas it can be deemed of lesser importance for C1.

\subsection{Localization} 
The \emph{aggregate bandwidth} 
plays a role in terms of delay/distance resolution and accuracy. For L1, a cluttered environment and high accuracy demand a bandwidth on the order of $\unit[2\text{--}4]{GHz}$ if the resolution is to be achieved in the delay domain, while it can go down to $\unit[500]{MHz}$ when resolution is achieved in the angular domain (see later in Section \ref{sec:HWALoc}). For L2, a bandwidth of $0.4\text{--}\unit[1]{GHz}$ is sufficient. {To meet the L2 update rate requirement, the \ac{E2E} latency should be below $\unit[1]{ms}$, which can be met with the considered bandwidth, though $\unit[1]{GHz}$ is preferred.\footnote{Considering OFDM-like waveforms with 4096 subcarriers and 7\% cyclic prefix overhead, OFDM symbol duration are approximately $\unit[1\text{--}10]{us}$ for bandwidths in $\unit[0.4\text{--}4]{GHz}$. Considering slots of 14 symbols, physical layer latency with bi-directional transmission is on the order of $\unit[28\text{--}280]{us}$.}} Under L3, a bandwidth
of less than $\unit[400]{MHz}$ is sufficient. The \emph{waveform type} is largely irrelevant for localization, provided it can be flexible in duration and has a suitable ambiguity function with narrow main lobe and suppressed sidelobes. In terms of \emph{modulation type}, since localization is based on pilots, low-order constant-modulus signals are preferred for L1--L3. 
Communication should be coherent for L1 and L2 to achieve accuracy beyond the resolution limit, while L3 can rely on more simple non-coherent measurements. 
To optimize tracking performance, \emph{signal shaping} should be employed in L1 and L2, both for delay estimation (shaping in frequency, leading to a preference for OFDM-like signals) \cite{OFDM_DFRC_TSP_2021} and angle estimation (shaping in beamspace) \cite{CRB_JRC_Beamforming_TSP_2021}. 

\subsection{Sensing} 
To support cm-level ranging accuracy and resolution in S1 and S2, the \emph{aggregate bandwidth} should be on the order of $\unit[2\text{--}4]{GHz}$.
Similar to localization, the \emph{waveform type} affects the sensing performance only through the main-lobe and side-lobe characteristics of the corresponding ambiguity function. Multi-carrier waveforms, such as \ac{OFDM} and \ac{OTFS}, have the advantages of lower side-lobe levels \cite{IEEE80211ad_radar_TWC_2020} and greater flexibility in signal shaping \cite{General_Multicarrier_Radar_TSP_2016} and removal of the transmit data, over single-carrier ones, at the cost of being less hardware-friendly. With regard to the \emph{modulation type}, sensing requires constant-modulus signals to have favorable side-lobe behavior. S1 relies on both random data and pilots (constant-modulus), while S2 uses only pilots. Hence, S1 can be more sensitive to the modulation type than S2. To achieve high range and angular accuracy in S1 and S2, both use cases should employ coherent measurements. Finally, for the \emph{signal shaping}, pilot allocation could be crucial for S2, while for S1, there are trade-offs between communication rate and sensing performance~\cite{OFDM_DFRC_TSP_2021}

\section{Requirements on Hardware Architectures}
The larger bandwidths identified in Section \ref{sec:Signals} are available 
at the upper \ac{mmWave} and terahertz frequency range. 
High propagation loss per patch type antenna element at these frequencies drastically limits the signal transmission distance and communication coverage range. In this case, \acp{TX} with high output power using antenna arrays is of particular importance, because the degradation of transmitted signal power due to high-frequency hardware technologies highly limits  performance. The key hardware impairments include nonlinear distortion (due to power amplifiers, leading also to intermodulation products outside the channel bandwidth), phase noise (induced by RF oscillators at the \ac{TX} and \ac{RX} chains), quantization noise and mutual coupling between antenna ports.

\begin{figure}[t]
\centering
\include{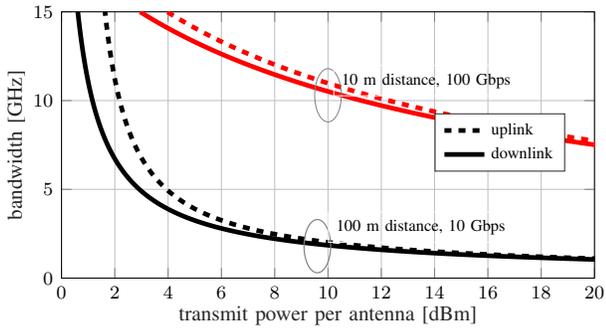}
\vspace{-5mm}
\caption{Example of required bandwidth vs required \ac{TX} power per antenna element for achieving a link distance of 10 m and 100 m. A point-to-point communication link with free-space propagation at 140 GHz was considered. Other parameters: 256 \ac{BS} antennas, 64  \ac{UE} antennas, \ac{BS} noise figure  5 dB, \ac{UE} noise figure is 10 dB, pathloss exponent  2. The sum of power amplifier output back off and \ac{RX} sensitivity degradation due to implementation imperfections is 20 dB. For details of the calculation, see \cite[Section~3.4.2.2.1]{Hexa-X-D2.1}.}
\label{fig:MurisFig}
\vspace{-3mm}
\end{figure}

\subsection{Communication} \label{sec_req_hw_comm}
In Fig.~\ref{fig:MurisFig} the tradeoff between bandwidth and output power for the  use-cases C1 and C2 is illustrated. For the simulation, a point-to-point communication link at a carrier frequency of 140~GHz was assumed. For C2, a bandwidth of around 2~GHz is sufficient, and that for both the uplink and downlink a saturated output power per antenna of around 5~dBm is needed. Increasing the power will not significantly reduce the bandwidth required. For realizing C1 at the same power levels, a bandwidth of around 13~GHz is needed. Note that for the uplink case the required power per element is larger since a smaller number of transmit elements is used compared to the downlink. To mitigate the need for a higher output power per element on the \ac{UE} side, it is possible to use more than one \ac{IN} in the uplink, \ac{D-MIMO} and perform multi-stream transmission. Achieving C1 with a bandwidth of 7~GHz and a single stream is very challenging due to the high output power that would be required, which is another argument for introducing multi-stream/multi-\ac{IN}/D-MIMO transmission. 

\subsection{Localization}  \label{sec:HWALoc}
L1 and L2 require large bandwidths, which is only available at a high \emph{carrier frequency}. \emph{Channelization} 
can be used, but 
phase coherence must be maintained across the entire band for coherent angle and delay estimation (and thus localization). 
For L3, lower bands (e.g., at $\unit[30]{GHz}$ or even $\unit[6]{GHz}$) are preferred with lower path loss. 
Localization relies on pilot signals, so that single-stream transmission is sufficient for broadcast signals. When dedicated pilots are used for individual devices or groups of devices, multi-stream pilots are preferred. Hence, in terms of \emph{array types}, planar analog and hybrid structures are sufficient for L1. If stringent latency is required as in L2, digital array structure can help with multi-streams and highly flexible pilot signal design to perform localization within fewer snapshots. Regarding the use case L3, single-antenna \acp{UE} are appropriate, due to lower hardware and computation cost. 
In addition to the beamforming gain, the array size at the BS determines the positioning accuracy, whereas the array size at the \ac{UE} affects the orientation estimation. The requirements on \emph{array size} are thus variable: for L3, 
a
single  omni-directional antenna  at the \ac{UE} and an array at the \acp{IN} can be sufficient. For L1 and L2, \acp{IN} array sizes depend on the SNR, position resolution, and accuracy requirements, as well as the considered bandwidth. If a large bandwidth is employed (L1 and L2), this relaxes requirements on the array sizes down to $10$--$20$ elements per dimension are sufficient, with an angular resolution of $\unit[5\text{--}10]{^{\circ} }$. With bandwidths below $\unit[500]{MHz}$, resolution in the angular domain requires on the order of $50\text{--}100$ elements per dimension in L1 ($\unit[1\text{--}2]{^{\circ} }$ resolution). High orientation accuracy demands large array sizes on the user-side, where $10\text{--}20$ elements per dimension should be considered. 
To understand the impact of the \emph{transmission power}, it is instructive to consider a simplified 2D uplink scenario (as in Fig.~\ref{fig:tradeoff}), where the localization accuracy in uplink can be expressed as \cite{abu2018error}:
\begin{align}
    \text{SPEB}[\text{m}^2]=\frac{ N_0 B d^2}{T P_{\text{tx}}\lambda^{2}}\left(\frac{c^{2}\alpha_{\text{range}}}{B^{2}N_{\text{rx}}}+\frac{d^{2}\alpha_{\text{angle}}}{N_{\text{rx}}^{3}}\right), \label{eq:SPEB}
\end{align}
where $T$ is the integration time (in number of transmissions),  $B$ is the aggregate bandwidth, $N_0$ the noise power spectral density, $d$ is the distance between \ac{UE} and \ac{IN}, $\lambda$ the wavelength, $c$ the speed of light, $N_{\text{rx}}$ the number of  \ac{IN} antennas, and $\alpha_{\text{range}}$, $\alpha_{\text{angle}}$ are constants. Finally, $P_{\text{tx}}$ is the transmit power. The first term in \eqref{eq:SPEB} captures the distance estimation error, while the second the angle estimation error. Hence, solving  \eqref{eq:SPEB} for  $P_{\text{tx}}$ yields a minimum transmission power for a certain target accuracy. We see that  $P_{\text{tx}}$ can be reduced by increased number of antennas $N_{\text{rx}}$  or longer integration times $T$. The latter approach places demands on the stability of oscillators. For this reason, L1 (operating at short ranges with relaxed integration times) does not require higher transmission power, while L2 and L3 do (the former due to the limited integration time, the latter due to the long link range).

\subsection{Sensing} 
Both S1 and S2 require operation at high \emph{carrier frequency} due to utilization of large bandwidths for high accuracy and resolution in range. From the perspective of frequency \emph{channelization} \cite{channelization_2017}, S1 and S2 should perform coherent processing across the entire bandwidth. 
In terms of \emph{array types}, monostatic sensing (S1) can exploit (re-use) both data and pilot symbols generated by the communication system (i.e., opportunistic sensing \cite{IEEE80211ad_radar_TWC_2020}); hence, the \ac{TX} array structure will mainly be determined by the communication requirements (see Sec.~\ref{sec_req_hw_comm}), while analog and hybrid arrays are sufficient for sensing purposes at the \ac{RX} side. On the other hand, bi-static sensing (S2) favor pilot symbols due to physically separate \ac{TXRX} hardware, which might degrade the performance of range estimation (availability of a smaller portion of time-frequency resources) and angle estimation (decrease in SNR) compared to S1. Hence, S2 \ac{RX} can be equipped with a hybrid or digital array to compensate for this performance loss. Moreover, to support $1^\circ$  degrees of angular resolution for S2, the \ac{RX} \emph{array size} should be at least 100 elements per dimension (35 for S1). Contrary to the \ac{RX} side, the array size at the \ac{TX} has no impact on the angular resolution. The array size can thus vary depending on requirements on the range and angle accuracy, which, in turn, are functions of SNR, bandwidth and \ac{RX} array size. In the case of large bandwidths ($\unit[2\text{--}4]{GHz}$) and/or large \ac{RX} arrays (100 elements per dimension), small \ac{TX} arrays with $10$--$20$ elements per dimension can be sufficient to support both S1 and S2 accuracy requirements, while for smaller bandwidths (below $\unit[1]{GHz}$) and smaller \ac{RX} arrays ($10$--$20$ elements per dimension), S1 and S2 requires $40$--$50$ elements per dimension at the \ac{TX} array to boost SNR towards the desired object locations.
An important distinction between localization and sensing pertains to the \emph{transmission power}. Since sensing (especially, S1) needs to combat higher path loss ($d^4$, where $d$ is the distance) than localization ($d^2$), the transmit power should be larger for sensing. However, such high power requirements for sensing may bring additional hardware complexities. For monostatic sensing (S1), the \ac{TX} and receiver are co-located and work in full-duplex mode, which necessitates perfect decoupling of \ac{TXRX} antenna arrays to prevent self-interference \cite{Interference_MIMO_OFDM_Radar_2018,80211_Radar_TVT_2018,barneto2019full}. On the other hand, for bi-static sensing (S2), the \ac{TXRX} arrays are far from each other, meaning that perfect isolation is not an issue.

\section{Requirements on Deployments}
\label{sec:deployments}
Seen from a joint communication, localization, and sensing performance perspective, the network planning problem is complex. The main reason is that multiple metrics need to be optimized simultaneously: coverage, capacity, positioning error, sensing error. 

\subsection{Communication}
Traditionally, the deployment of communication networks targeted {\em wide coverage} and {\em high capacity}~\cite{ClLoHoRaKu18} subject to cost constraints. Except for the compulsory emergency location services provided by the cellular network operators under regulatory mandates, localization and sensing services were disregarded in the network planning phase. By contrast, the future 6G networks shall provide integrated communication and sensing services to support emerging use cases such as industrial applications, augmented reality, etc. 
Moreover, fulfilling challenging joint coverage, reliability and throughput requirements can be made possible by distributed   deployments.
The promising \ac{D-MIMO} communication technology inherently implies (sub-)nanosecond synchronization among densely deployed network nodes, as well as multiple antennas both at \ac{TX} and \ac{RX}, which 
benefit from reduced path loss and macro-diversity mitigating shadowing/blocking. 

\subsection{Localization} \label{sec:DeploymentsLocalization}
Similar to communications, the \emph{placement} of localization infrastructure also targets {wide coverage}, but its primary goal is to minimize the {localization error} rather than rate. 
Also different from communication, localization of a device requires a plurality of \acp{IN} and the corresponding performance depends not only on the SNR, but also on the relative geometry of the \acp{IN}, through the so-called \ac{GDOP}: with delay-only measurements each device needs \ac{LoS} to at least 3 \acp{BS} in \ac{LoS} under \ac{RTT} and 4 under TDoA for \acp{UE} in the convex hull of these BSs. With angle-only measurements, at least 2 \acp{IN} (2 BSs or one BS and one \ac{RIS}, 
which is a minimal configuration). In addition to adopting multiple \acp{IN}, geometry diversity can be achieved by exploiting the multipath components~\cite{witrisal2016high}. In this case, the surrounding environment can serve as passive \acp{IN} (e.g., virtual anchors) for a better localization performance. While BS placement will be limited to existing sites or other constraints, \acp{RIS} can be placed in an optimized manner to meet GDOP requirements. 
Based on these considerations, L1 and L2 would need at least 2  \acp{IN} visible at any time, while L3 should rely on at least 3--4 \acp{IN}, as long-range links rely on time-based measurements (cf. Fig.~\ref{fig:tradeoff}). 
The \emph{synchronization} requirements are a function of the employed time-based measurements.\footnote{\Ac{AoA} measurements only need rough synchronization to know which BS precoder or \ac{RIS} configuration is used at which time.} 
\Ac{TDoA} requires fine timing synchronization among the \acp{IN}, on the order of $\unit[1\text{--}10]{ns}$ for L3, around $\unit[0.5]{ns}$ for L2, and around $\unit[100]{ps}$ for L1. For coherent combining across different \acp{IN} (e.g., using \ac{D-MIMO}), synchronization requirements are even more demanding, as they relate to the signal phase. To relieve these requirements, \ac{RTT} measurements can be used, which only impose local clock stability requirements, but scale worse than \ac{TDoA} under high user density.
Estimating the synchronization error (clock offset) during the localization phase is another option to mitigate the effect of clock drift.
Finally, accurate localization requires accurate \emph{knowledge of the locations} of the \acp{IN}: both the 3D location, the 3D orientation (for \ac{AoA}), and the phase center must be determined. L3 can cope with errors on the order of tens of centimeters. 
For L2 and L1, extreme demands are placed on calibration, which will likely require novel procedures (either offline or online). Since small orientation errors lead to large location errors (see Fig.~\ref{fig:tradeoff}), extremely precise orientation calibration is needed for 
\acp{IN}, when \acp{UE} are farther away from the \acp{IN}. 

\subsection{Sensing}
In monostatic sensing (S1), the \emph{placement} should be such that it provides sufficient coverage in the deployment region. Multiple \acp{IN} need to coordinate transmissions to avoid interference.
For bi- and multistatic sensing (S2), the placement problem is similar to localization (see Section \ref{sec:DeploymentsLocalization} with related \ac{GDOP} notions), though with larger path loss. The combination of angle and TDoA (with respect to the \ac{LoS} path) measurements provides a 3D picture of the environment.  Statistical error bounds may be used to optimize the deployment jointly for localization and sensing in both S1 and S2. 
In terms of \emph{synchronization}, S1 does not impose any requirement, while under S2, tight inter-\ac{IN} synchronization is needed when no \ac{LoS} link is present between the \ac{TX} and \acp{RX}. Finally, in terms of \emph{\ac{IN} knowledge}, S1 imposes no requirements, as sensing is performed in the frame of reference of the monostatic sensor. For S2, requirements are similar to localization.

\section{Conclusions}
\label{sec:conclusion}
6G use cases feature extreme requirements in terms of communication, localization, and sensing \acp{KPI}. The purpose of this paper was to systematically investigate the corresponding requirements in signals, hardware architectures, and deployments. From the summarizing Table \ref{tab:SummarizingTable}, it is apparent that to support a combination of use cases, the following combination is preferred: aggregate bandwidths around $5\text{--}10$ GHz with channels of around $2$ GHz at carriers above 100 GHz, flexible multi-carrier waveforms and coherent communication, with possibility to shape signals in space, relying on analog or hybrid arrays with tens of elements per dimension at both transmitter and receiver side. Transmit power requirements are moderate to high, but each user must be able to have \ac{LoS} visibility to several \acp{IN}, which may include \acp{RIS}. Very high demands are placed on calibration in terms of location and orientation knowledge of the \acp{IN}, as well the synchronization.

\section*{Acknowledgment}
This work was supported, in part, by the European Commission through the H2020 project Hexa-X (Grant Agreement no. 101015956) and the MSCA-IF grant
888913 (OTFS-RADCOM).

\balance
\bibliographystyle{IEEEtran}
\bibliography{IEEEabrv,bib/hexa_x_vision}
\end{document}